\documentclass[]{aastex631}
\usepackage{url}
\usepackage{xcolor}
\usepackage{CJK}
\usepackage{amsmath,amssymb}

\shorttitle{A Transition Discovered in the Subcritical Regime of 1A 0535+262}
\shortauthors{Xiao et al.}

\graphicspath{{./}{figures/}}

\begin{document}

\title{A Transition Discovered in the Subcritical Regime of 1A 0535+262}

\author{Hua Xiao}
\affiliation{School of Physics and Astronomy, Sun Yat-sen University, Zhuhai, 519082, People's Republic of China}

\author[0000-0001-9599-7285]{Long Ji\textsuperscript{*}}
\email{jilong@mail.sysu.edu.cn}
\affiliation{School of Physics and Astronomy, Sun Yat-sen University, Zhuhai, 519082, People's Republic of China}

\begin{abstract}
We present {\it NICER} observations of accreting X-ray pulsar 1A 0535+262 during its faint state ($\lesssim 6\times10^{36}$\,erg/s) observed in several type-I and II outbursts. 
We discovered a transition of temporal and spectral properties around the luminosity $L_{\rm t}=3.3\times10^{35}$\,erg/s, below which spectra are relatively soft and the pulse profiles have only a narrow peak.
The spectra are harder and a secondary hump gradually appears in pulse profiles when $L \gtrsim L_{\rm t}$.
We discuss possible physical mechanisms for this transition, including different Comptonization seed photons, the disappearance of gas shocks on the neutron star surface, and the combination of plasma and vacuum polarization effects.
\end{abstract}


\section{Introduction \label{sec:intro}}

X-ray pulsars (XRPs) consist of a highly magnetized neutron star and an early-type (O/B) companion star, powered by the accretion process between binary systems \citep[for a review, see][]{Mushtukov2022}. The neutron star usefully exhibits a strong magnetic field of $\sim10^{12–13}$\,G.
Therefore, around its magnetosphere the accreted material couples to magnetic field lines and is channeled onto polar caps of the neutron star, resulting in X-ray pulsations.
During the last few decades, extensive theoretical studies have been done on the accreting structure in this vicinity of polar caps, and two classical accretion regimes have been proposed, depending on the accretion rate \citep[][]{Basko1975, Basko1976, Burnard1991, Postnov2015, Becker2012, Mushtukov2015, Becker2022}. 
In the supercritical regime, when $L \gtrsim1.5\times10^{37}B_{12}^{16/15}$\,erg/s (where $L$ is the luminosity and $B_{12}$ is the magnetic field in units of $10^{12}$\,G), the accreted matter is decelerated by a radiation-dominated shock and an accretion column at some distance above the surface of the neutron star. In this case, most of the radiation diffuses within the accretion column and escapes through the column wall, forming the so-called ``fan beam" \citep[e.g.,][]{Becker2012}.
On the other hand, for very faint states ($L \lesssim 10^{34-35}$\,erg/s), the accreting matter falls onto the surface of the neutron star in near-free fall and is prevented by Coulomb interactions \citep{Sokolova-Lapa2021, Mushtukov2021}. The radiation is produced by hot spots or mounds near polar caps and propagates mainly along the magnetic field lines, forming the so-called ``pencil beam". 
Between these two extremes, the situation is more complex, where the emission may be a hybrid combination of ``pencil" and ``fan" patterns \citep{Blum2000, Hu2023}.
In addition, collisionless shocks for $L \lesssim 10^{36}$\,erg/s were proposed, although the underlying physics is still unknown \citep{Langer1982, Bykov2004}.

Observations dedicated to the faint state are relatively rare. 
Based on \textit{NuSTAR} observations, \citet{Tsygankov2019a} reported a obvious spectral evolution in GX 304-1, from the exponential cutoff powerlaw shape at $L\sim 10^{36-37}$\,erg/s to a double hump shape at a lower luminosity $L \sim 10^{34}$\,erg/s.
This two-hump shape was also observed in other faint sources, e.g., 1A 0525+262 and X Persei \citep{Tsygankov2019b, Doroshenko2012}.
It is believed that the ``thermal" component at low energies originates from the deep layer of the neutron star atmosphere, and the high-energy hump can be explained as the resonant Comptonization in the heated non-isothermal part of the atmosphere and cyclotron photons \citep{Sokolova-Lapa2021}.

Discovered by Ariel V in 1975 \citep{Rosenberg1975}, 1A 0535+262 is a transient X-ray accreting pulsar with a period of $\sim104$\,s \citep{Rosenberg1975} and a distance of $\sim2$\,kpc \citep{Steele1998}. Its orbital period and eccentricity are 110.3d and 0.47, respectively. 
In recent years, this source experienced several type-I (normal) outbursts and a type-II (giant) outburst in 2020. Type-I outbursts are normally associated with periastron passages, and have a lower peak luminosity ($\leq10^{37} \rm erg/s$) compared with the giant outburst and are suitable for studies in the low accretion rate state. 
In this source, \citet{Tsygankov2019b} revealed the presence of two spectral components at $L\sim7 \times 10^{34}$\,erg/s,
while \citet{Ballhausen2017} suggested that the spectrum can be well described as a single component continuum model at $L\sim2\times 10^{36}$\,erg/s, and strong residuals appear around 30\,keV at $L\sim0.6\times 10^{36}$\,erg/s although they are model-dependent.
This implies that a smooth evolution may occur, bridging between the two spectral shapes, and thus motivates us to perform a detailed analysis around $L\sim10^{36}$\,erg/s.
In this paper, we present the timing and spectral analysis of 1A 0535+262 during its faint state observed with \textit{NICER}.
This paper is structured as follows. In Section \ref{sec2}, we describe details of the observations and data reduction; the results are presented in Section \ref{sec3}; and in Section \ref{sec4} we discuss the observational findings and their implications. 

\section{Observation and Data Reduction \label{sec2}}
\begin{figure}[ht!]
	\includegraphics[width=0.8\linewidth]{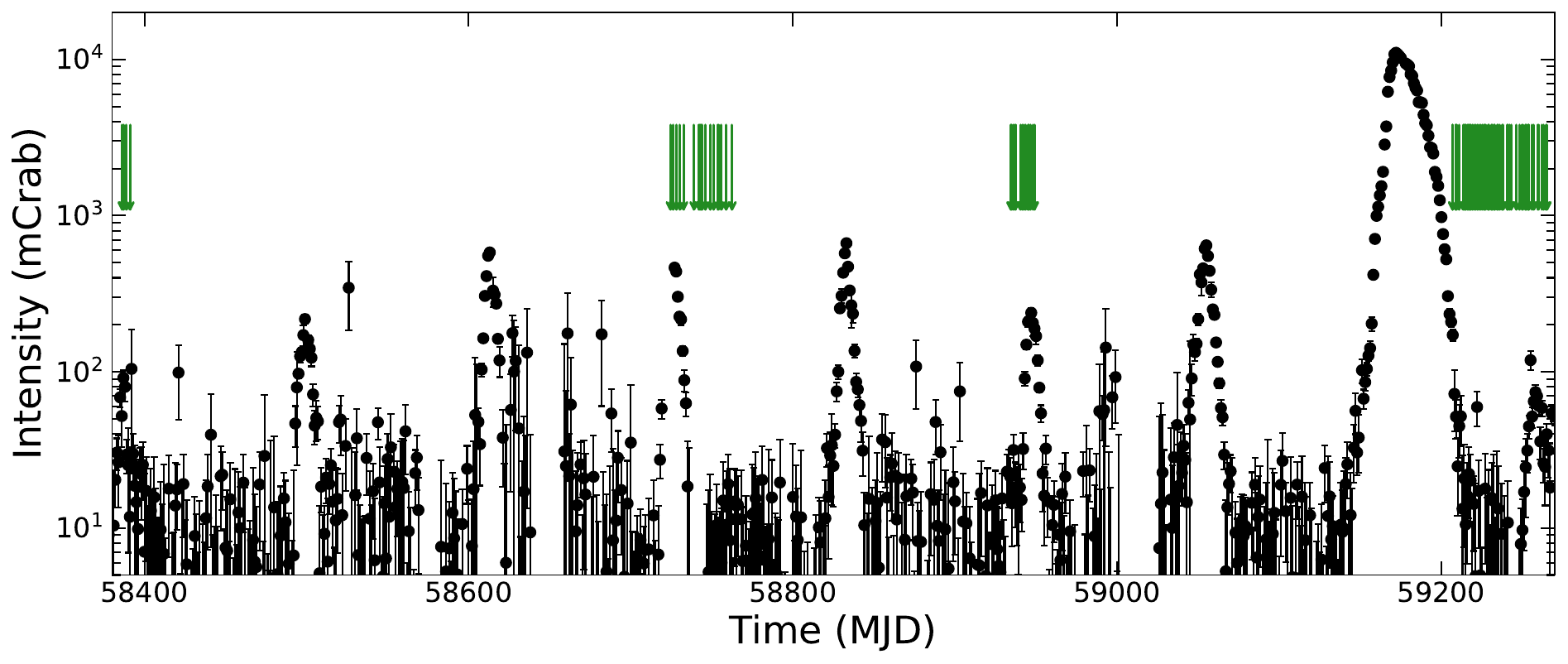}
	\centering
	\caption{Long-term lightcurve of 1A 0535+262 observed with \textit{Swift}/BAT in the energy range of 15-50\,keV.
    This source experienced several type-I outbursts and a giant outburst around MJD 59180.
    {\it NICER} observations used in this paper are marked with green arrows. 
	\label{fig:1}}
\end{figure}

The \textit{Neutron star Interior Composition Explorer} ({\it NICER}) is a payload onboard the International Space Station, dedicated to timing and spectroscopic analysis at soft X-rays (0.2-12\,keV) \citep{Gendreau2016}. 
Its high sensitivity makes it possible to study transient sources even in their faint states.
Over the past few years, 1A 0535+262 experienced several type-I outbursts and a giant type-II outburst in 2020-2021.
Since the bright state of the giant outburst has been extensively studied by different authors in different aspects \citep{Kong2021, Kong2022, Wang2022, Mandal2022, Reig2022, Hu2023, Chhotaray2023, Liu2023}, it is excluded from our sample.
In this paper, we focus only on 82 pointing observations from type-I outbursts and the faint state of the giant outburst, as shown in Figure~\ref{fig:1}.
%

%
The data reduction was performed by using the official software {\sc heasoft v6.31.1}/{\sc nicerdas} package (version {\tt \string 2022-12-16\_V010a}) and the latest calibration database ({\sc caldb}, version {\tt \string xti20221001}).  
We used the tool {\tt nicerl2}\footnote{\url{https://heasarc.gsfc.nasa.gov/docs/nicer/nicer_analysis.html}} to screen clean events, and extracted light curves and spectra with {\tt nicerl3-lc} and {\tt nicerl3-spect}, respectively.
We estimated the background by using the {\it scorpeon}\footnote{\url{https://heasarc.gsfc.nasa.gov/docs/nicer/analysis_threads/scorpeon-overview/}} model.
During the timing analysis, the barycentric correction was performed by using the tool {\tt barycorr}.
We adopted the epoch-folding method ({\tt efsearch}) to estimate the spin period of the source and folded its pulse profiles using the tool {\tt efold}.
The spectral analysis was performed by using the X-ray spectral fitting package ({\sc xspec}) version 12.13.0c \citep{Arnaud1996}. 
All uncertainties in this paper correspond to a confidence level of 68\%.

%
%

\newpage
\section{Results} \label{sec3}
\subsection{Spectral analysis} \label{sec3.1}
\begin{figure}[ht!]
	\includegraphics[width=0.45\linewidth]{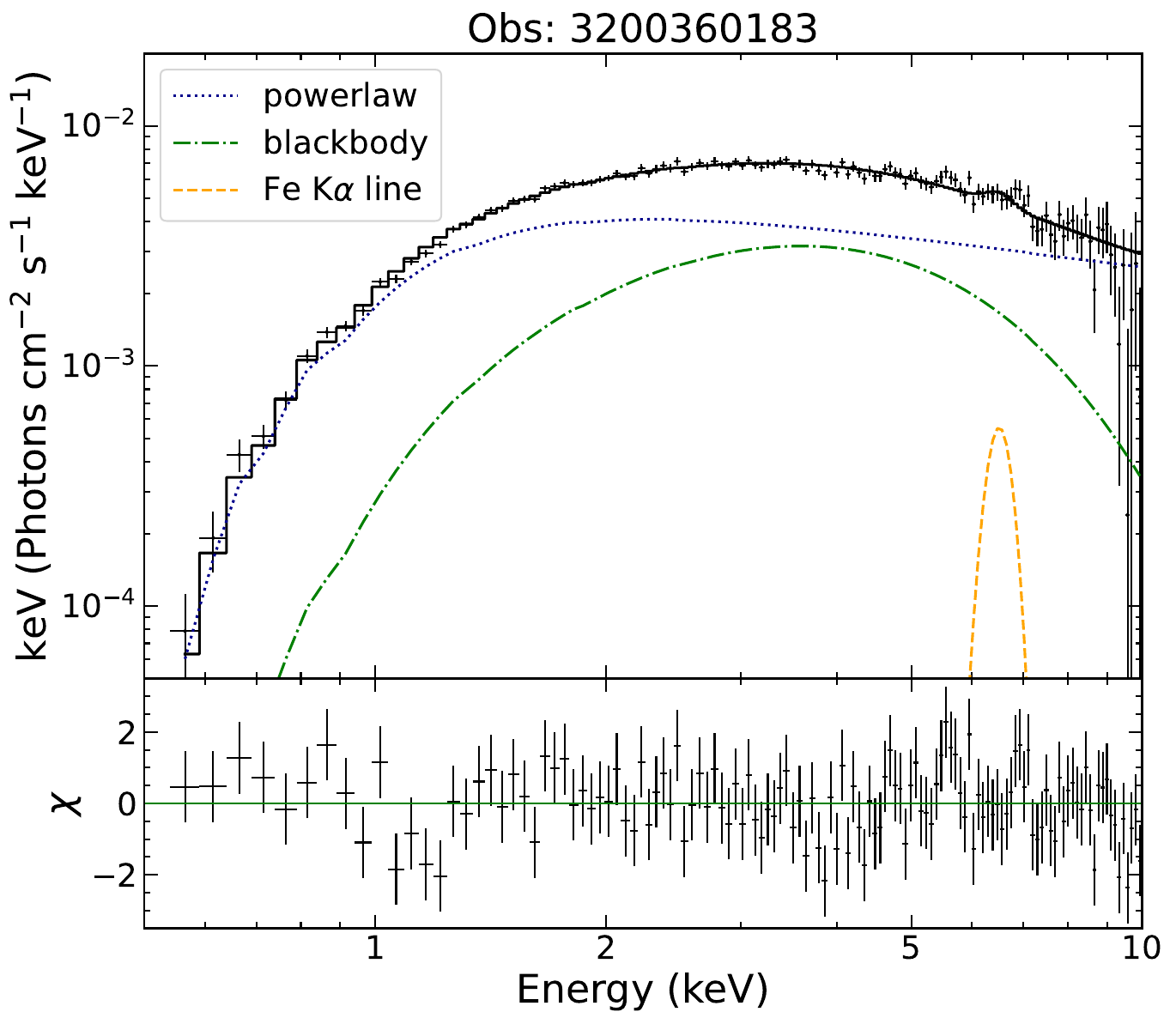}
    \includegraphics[width=0.45\linewidth]{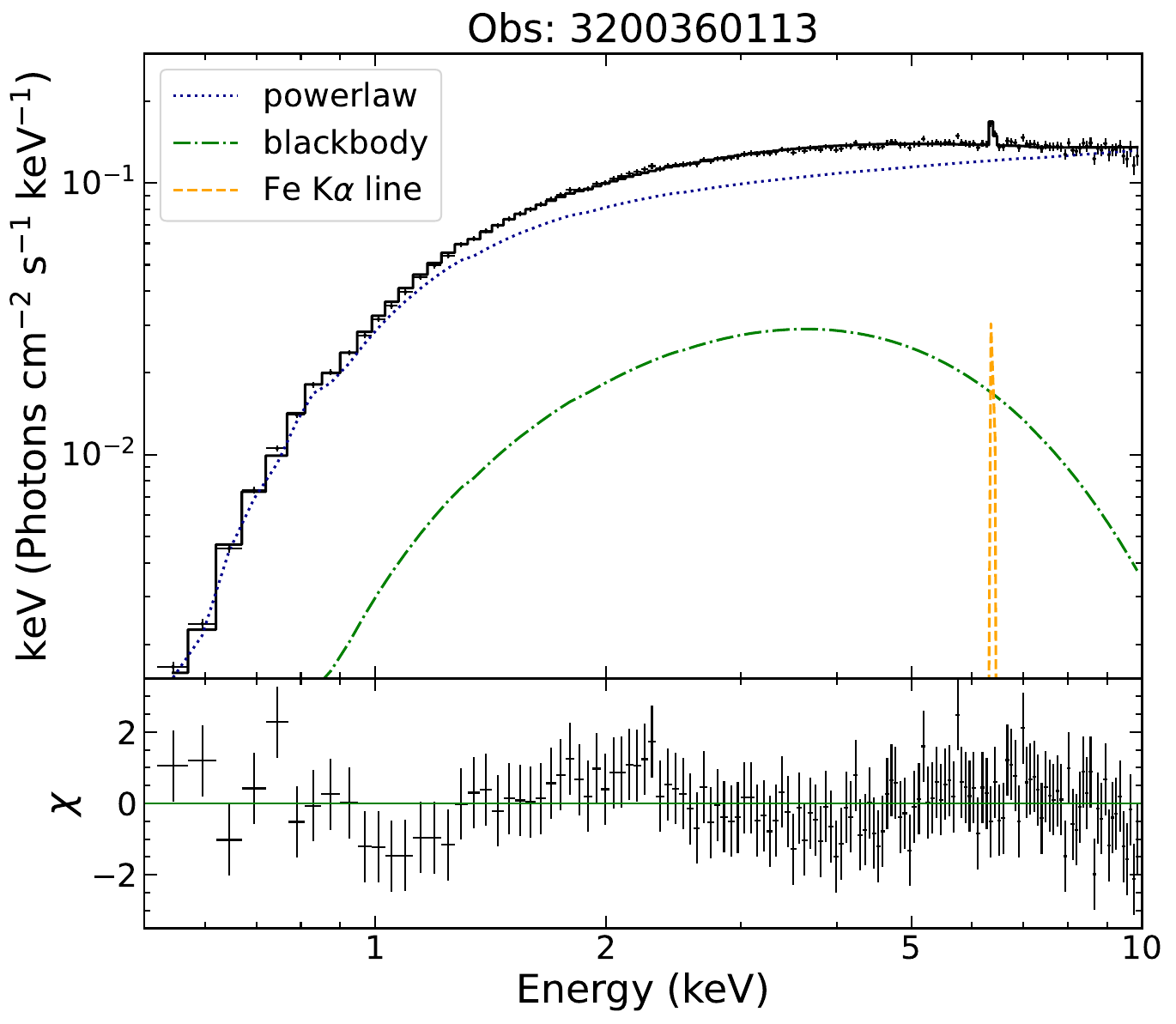}
	\centering
	\caption{Representative spectra of 1A 0535+262 at luminosities of $1.1 \times 10^{35}$\,erg/s (left) and $2.7 \times 10^{36}$\,erg/s (right), respectively.
	\label{fig:2}}
\end{figure}

\begin{figure}[ht!]
	\includegraphics[width=0.85\linewidth]{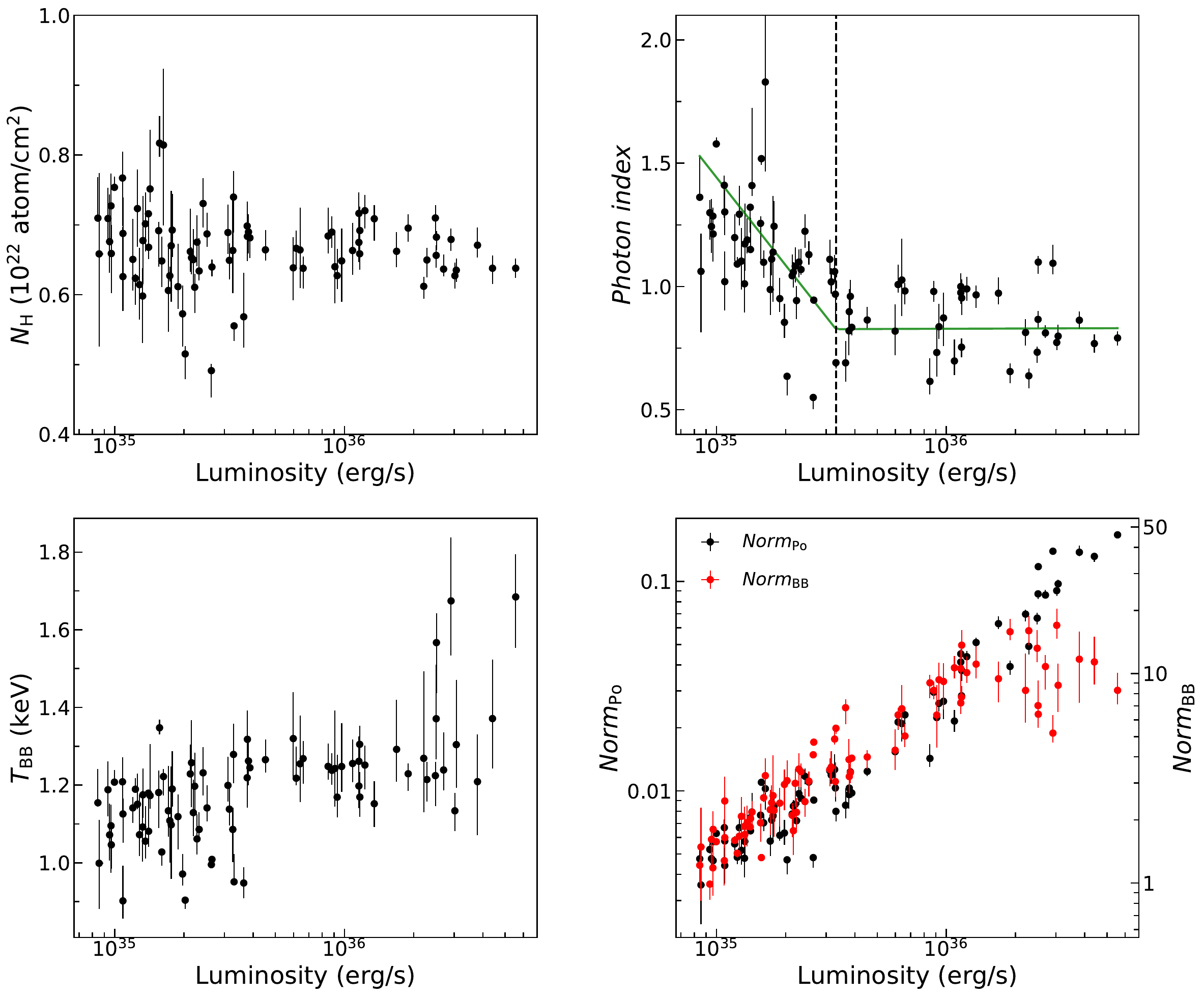}
	\centering
	\caption{
	The evolution of spectral parameters in the faint states of 1A 0535+262 using the model of {\tt Tbabs*(powerlaw + bbodyrad + gauss)}. 
    The green line shows the evolution of the photon index modelled with a broken powerlaw function.
    The black dashed lines represent the transitional luminosity $L_{\rm t}$ at $\rm \sim3.3 \times 10^{35}\,erg/s$.
    }
	\label{fig:3}
\end{figure}

%
According to previous studies, the broadband spectrum of 1A 0535+262 can be described by phenomenological models such as a power-law component with a high energy cutoff ({\tt cutoffpl}) combined with a blackbody component, and an absorption line at $\sim$45\,keV caused by cyclotron resonance scattering features \citep[e.g.,][]{Caballero2013, Ballhausen2017, Kong2021}.
%
However, limited by {\it NICER}'s narrow band (0.5-10\,keV), complex models cannot be constrained.
Therefore, we fitted all the spectra with a similar but simplified model, consisting of a powerlaw and a blackbody component. In addition, a narrow Gaussian component was added to account for the Fe K$\alpha$ fluorescence line. 
We also used the {\tt Tbabs} \citep{Wilms2000} model to described interstellar absorption, i.e., the fitting model in {\sc xspec} is {\tt Tbabs*(powerlaw + bbodyrad + gauss)}.
This model was able to fit the data well (Table \ref{data}).
We note that our aim is not to find a physical model, but to select a phenomenological model that can be used to describe the spectral shape.
The evolution of spectral parameters is shown in Figure~\ref{fig:3}.
%
%
%
In general, both blackbody and powerlaw components increase with luminosity, but the latter increases more significantly.
In Figure~\ref{fig:2} we present two representative spectral shapes for low and high luminosities.
For a luminosity of $\sim10^{35}\,$$\rm erg/s$, two continuum components are comparable, while the powerlaw component is dominating for the brighter state. 
The luminosity was calculated as $4\pi\,C\,F_{\rm 0.5-10}d^2$ assuming an isotropic radiation, where $d$=2\,kpc is the distance, $F_{\rm 0.5-10}$ is the flux at 0.5-10\,keV obtained from spectral fits, and $C=3.07$\footnote{In practice, the bolometric correction was estimated from the flux ratio between 0.1-100\,keV and 0.5-10\,keV when using a cutoff powerlaw model to fit the broadband {\it NuSTAR} observation (80001016004), when the luminosity is about $10^{36}$erg/s \citep{Ballhausen2017}.} 
is the bolometric correction.   
The equivalent hydrogen column density ($N_{\rm H}$) and the blackbody temperature ($T_{\rm bb}$) 
generally keep unchanged. 
On the other hand, the photon index ($\Gamma$) decreases with the increasing luminosity ($L$) at low luminosities, 
indicating a spectral hardening, 
while the slope seems to be lower at higher luminosities.
%
%
To describe the $L-\Gamma$ relation quantitatively, we fitted them with a powerlaw model and a broken powerlaw model, respectively.
For the latter, the resulting transitional luminosity ($L_{\rm t}$) is $(3.3\pm0.1) \times 10^{35}\,\rm erg/s$.
We note that the scattering of $\Gamma$ is dominated by systematic errors.
In practice, we assumed a same systematic error ($\sigma_{\rm sys}$) for all the points, and estimated its value $\sigma_{\rm sys}=0.12$ by making the reduced-$\chi^2$=1 when fitting the points of $L>10^{36}\,$$\rm erg/s$ with a constant line.
After considering the systematic error, the goodness-of-fits of the powerlaw model and the broken powerlaw model were $\chi^2$=124.42\,(80\,dof) and $\chi^2$=111.02\,(78\,dof).
We adopted the F-test for the model comparison
and found a probability of chance improvement of 0.01 for the latter model.
%
%
%

\subsection{Luminosity-dependent pulse profiles}
\begin{figure}[ht!]
	\includegraphics[width=0.9\linewidth]{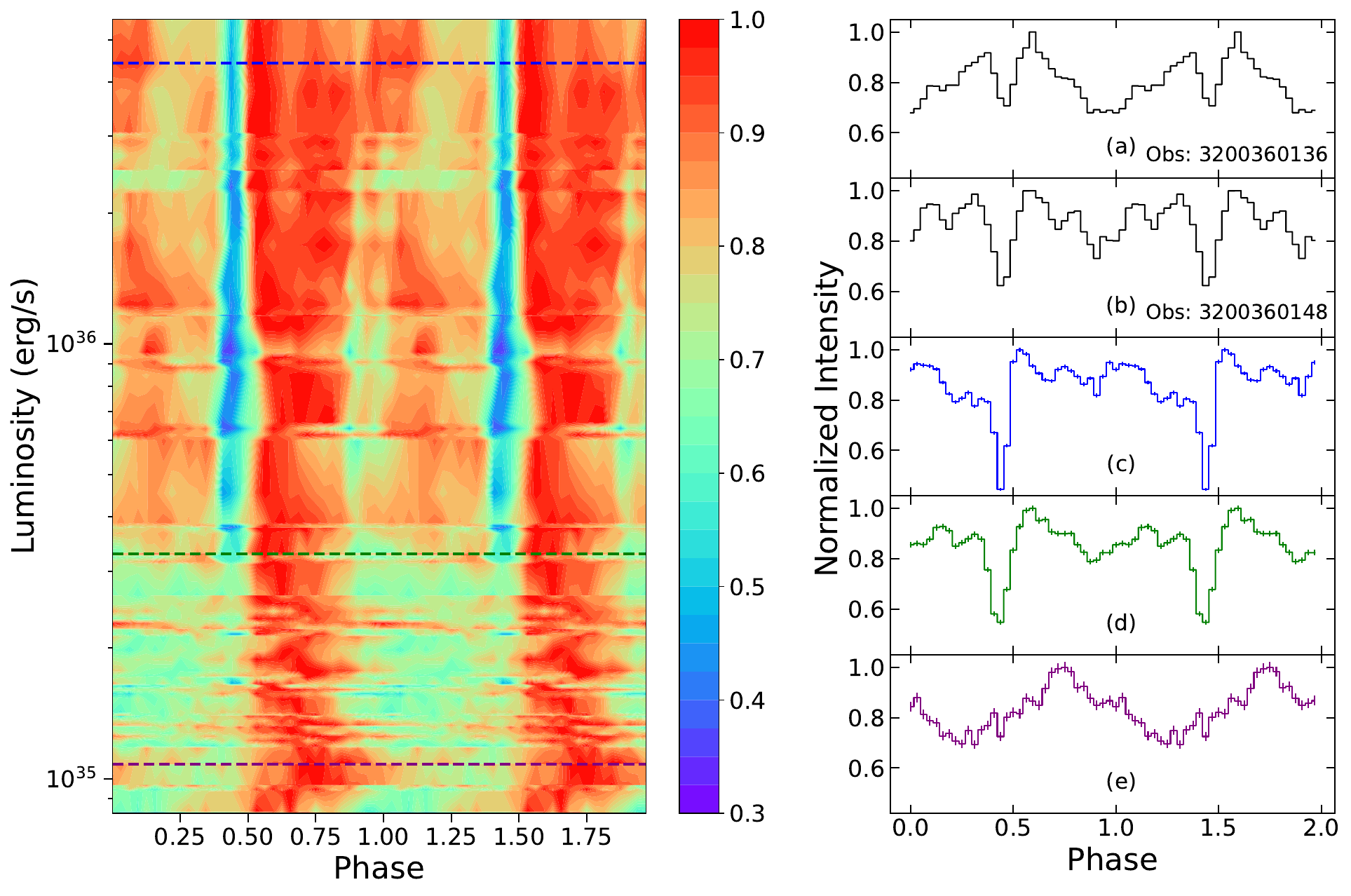}
	\centering
	\caption{Left: the evolution of pulse profiles with luminosity.
        For clarity, we normalized the pulse profiles by their peak values.
        Right: representative pulse profiles at luminosities of $4.4 \times 10^{36}$\,erg/s (panel c), $3.3 \times 10^{35}$\,erg/s (panel d) and $1.2 \times 10^{35}$\,erg/s (panel e), which are marked in the left panel with dashed lines. 
        For the sake of comparison, we also present two pulse profiles (panels a and b) observed in the brighter states with luminosities of $8.0 \times 10^{37}\,$erg/s and $4.3 \times 10^{37}\,$erg/s, respectively.
        }
		\label{fig:4}
\end{figure}

\begin{figure}[ht!]
	\includegraphics[width=0.5\linewidth]{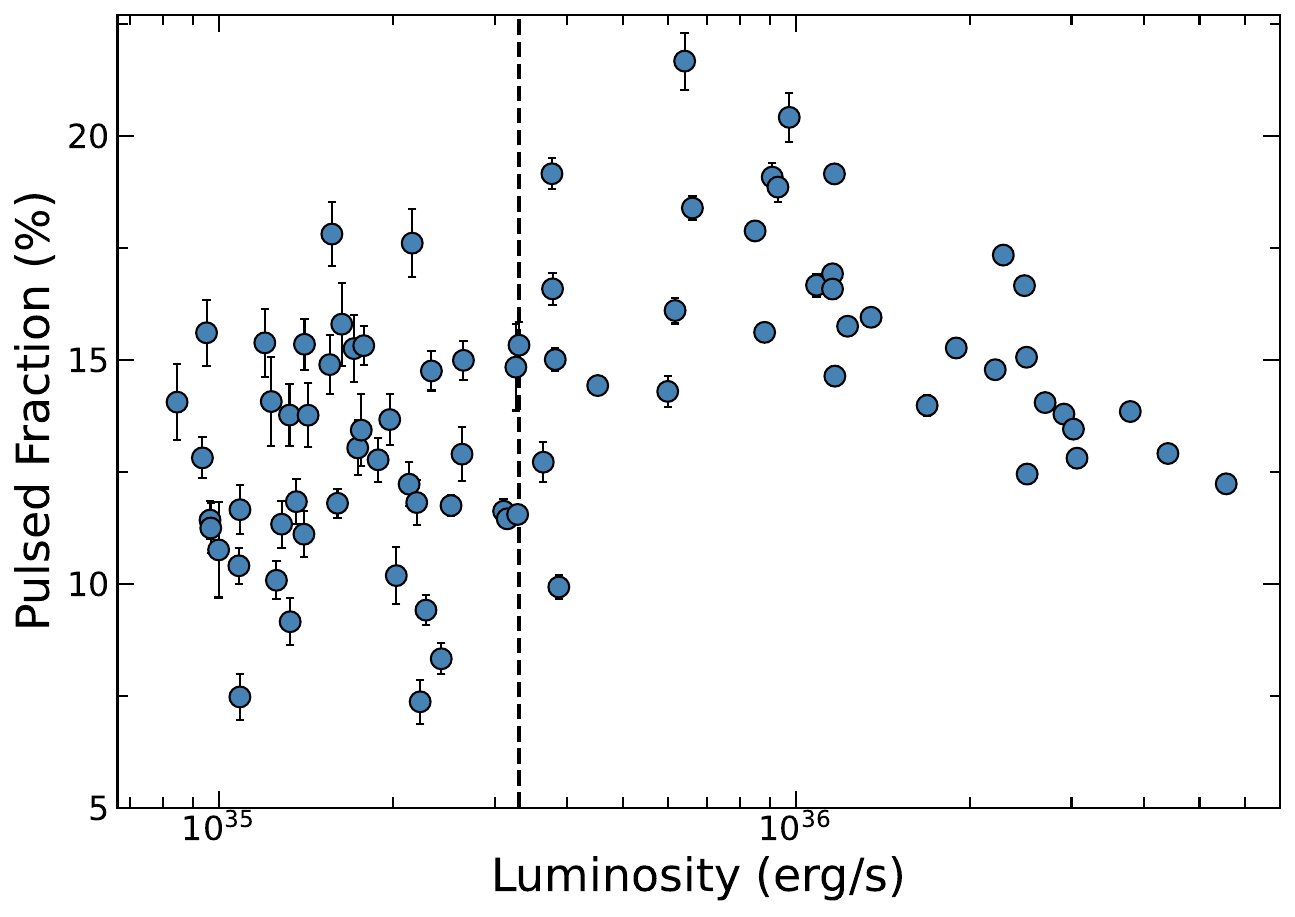}
	\centering
	\caption{Evolutions of pulsed fraction ($PF$) with luminosity for different \textit{NICER} observations. The black dashed line indicates the transitional luminosity obtained from the spectral analysis.
		\label{fig:5}}
\end{figure}

\begin{figure}[ht!]
    \includegraphics[width=0.9\linewidth]{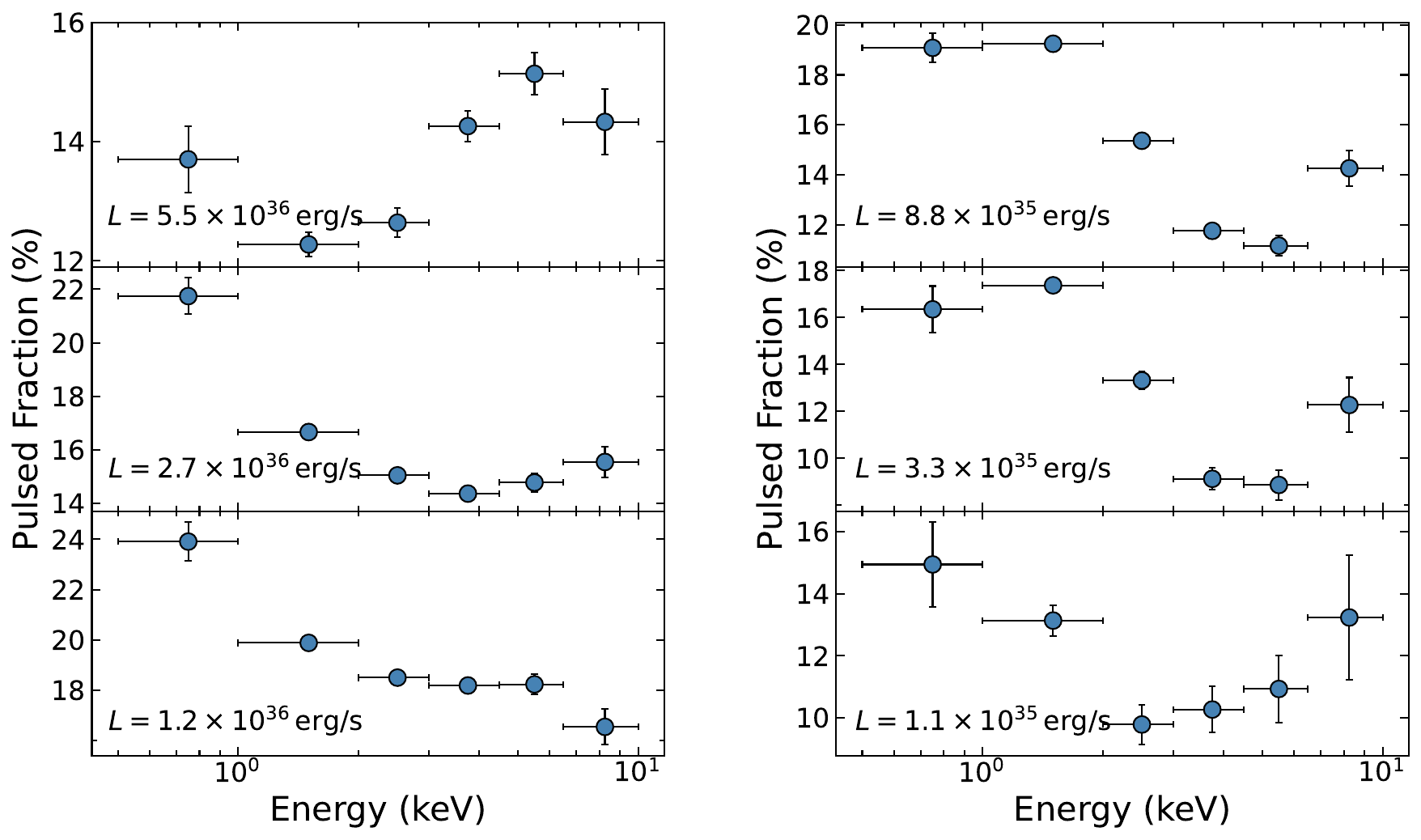}
	\centering
	\caption{Energy-dependent $PF$ at different luminosities.
		\label{fig:6}}
\end{figure}

%
%
%
%
%
%
%
%
%
%
%
For each {\it NICER} observation we searched for its periodicity by using the epoch-folding method.
Then we folded 1\,s-binned background-subtracted lightcurves to extract the pulse profiles.
We found that pulse profiles exhibit different patterns.
To investigate the luminosity-dependent evolution, we constructed a color-coded two-dimensional matrix in Figure~\ref{fig:4}, of which each row presents a pulse profile for a given luminosity.
For clarity, we aligned the pulse profiles using the cross-correlation method.
Here we used data taken from all available type-I outbursts and the faint state of the giant outburst, so that the detailed evolution of pulse profiles with luminosity can be well tracked. 
%
%
%
%
It is clear that the pulse profile exhibits a narrow peak at low luminosities, while around $3\times10^{35}$\,erg/s (i.e., the luminosity where the spectral shape changes), a narrow dip appears at the phase 0.5, accompanied with a secondary hump at the phase 0-0.4.
The hump increases at higher luminosities and eventually has a comparable intensity to the main peak (panel c in Figure~\ref{fig:4}).
For comparison, during the peak of the giant outburst, pulse profiles evolve into a double-peaked structure (panels a and b), which have been extensively studied by \citet{Wang2022, Mandal2022}.
In the right panel of Figure~\ref{fig:4}, we present representative pulse profiles from low to high luminosities.
%
In order to depict the pulse profile evolution quantitatively, we calculated the fractional root-mean-square (RMS) pulsed fraction ($PF$), which is defined as
\begin{equation}
f_{\rm rms} = \frac{( \textstyle \sum_{i=1}^{N}(r_i - \bar{r})^2 /N )^{1/2}}{\bar{r}} ,
\end{equation}
where $N$=32 is the number of phase bins, $r_i$ is the count rate in the phase bin $i$ and $\bar{r}$ is the phase-averaged count rate. 
The $PF$ errors were estimated using Monte Carlo simulations with $10^{4}$ samplings. 
The results are shown in Figure~\ref{fig:5}, where the $PF-L$ relation shows a large scattering for $L < L_{\rm t}$, while for $L > L_{\rm t}$ there is a significant negative correlation.
A sudden increase of the $PF$ may appear around $L_{\rm t}$, corresponding to the dramatic variation depicted in Figure~\ref{fig:4}.
Furthermore, we studied the energy-dependent $PF$ at different luminosities (Figure~\ref{fig:6}).
The $PF$ shows both negative and positive correlations with energy in 0.5-10\,keV.
The turn-over point ($E_{\rm p}$) between negative and positive correlations (i.e., the energy having a minimum $PF$) varies with luminosity.
For example, the $E_{\rm p}$ is around 2-3\,keV for the faint state ($L\sim10^{35}$\,erg/s), which is much lower than the results (i.e., $\sim$10\,keV) observed with {\it Insight-HXMT}, {\it NuSTAR} and {\it AstroSat} \citep{Wang2022,Mandal2022,Chhotaray2023} when the source is in brighter states.
%
%
%
%
%
%
%
%

\section{discussion} \label{sec4}
Using \textit{NICER} observations, we studied the temporal and spectral properties of accreting X-ray pulsar 1A 0535+262 during its faint state with a luminosity range of  $(0.8-56)\times10^{35}$\,erg/s in 0.1-100\,keV. 
The critical luminosity, corresponding to the onset of the accretion column, has been reported to be $\sim6\times10^{37}$\,erg/s based on the variations of cyclotron line energies \citep{Kong2021}.
Therefore, the results presented in this paper are concluded under the sub-critical accretion regime.
We found that the 0.5-10\,keV spectra can be well described as a simple model, i.e., the combination of a blackbody component and a powerlaw component. 
The spectral parameters are luminosity-dependent and a transitional luminosity $L_{\rm t}=3.3 \times 10^{35}$\,erg/s
is clearly discovered (Figure~\ref{fig:3}), which corresponds to an accretion rate of $ \dot{M}=2\times 10^{15}$\,g/s assuming the neutron star of a $1.4M_{\odot}$ mass and a $10^6$\,cm radius.
For $L\lesssim L_{\rm t}$, the $\Gamma-L$ relation is steeper and the relative contribution from the blackbody component is more significant, suggesting that the spectra are softer at low luminosities and are consistent with previous reports in GX 304-1, GRO J1008-57 and 1A 0535+262 \citep{Tsygankov2019a,Lutovinov2021,Tsygankov2019b}.
The pulse profiles also show a transitional evolution around $L_{\rm t}$, from a narrow peak at low luminosities to a borad peak at high luminosities.
This difference can be demonstrated in the pulsed fraction ($PF$) (Figure~\ref{fig:5}), where the $PF$ is relatively lower at low luminosities and the negative $PF-L$ is only shown at high luminosities.

Theoretical models of the accretion structure have been made by many authors assuming different considerations \citep{Langer1982, Becker2012, Mushtukov2021, Sokolova-Lapa2021, Becker2022}.
Generally, it is believed that the spectrum originates from Comptonization processes with seed photons from blackbody, bremsstrahlung and cyclotron emissions.
In soft X-rays, the blackbody is expected to be the dominant component of seed photons when the accretion rate is low, while the bremsstrahlung is the primary source at high accretion rates \citep[see Figure 7 and 9 in][]{Becker2022}.
Thus, the variable combination of seed photons can naturally interpret the spectral evolution we observed.
And in this case changes of pulse profiles can also be explained since the beam patterns of different Comptonization processes are different.
However, current models do not quantitatively predict at what luminosity the contributions of the two seed photons are comparable.
To verify this picture, detailed theoretical calculations around the luminosity $L_{\rm t}$ are required to compare with broadband spectral observations.

The observed phenomenon may be caused by other possibilities.
For example, as suggested by \citet{Langer1982}, the final deceleration of the matter will occur via a gas-mediated collisionless shock near the surface of the neutron star, which may lead to changes of timing and spectral properties.
This scenario was proposed to interpret the luminosity-dependent cyclotron line energy in Cepheus X-4 \citep{Vybornov2017}.
But we note that in 1A 0535+262 the cyclotron line energy keeps constant at low accretion rates \citep{Tsygankov2019a}, which makes the existence of the collisionless shock unclear.
In any case, if it is real, the transition from the Coulomb stopping (at higher luminosities) to the collisionless shock will happen around $ L_{\rm coul} \approx 1.17 \times 10^{37} B_{12}^{-1/3}$\,erg/s \citep[see Figure~1 in][]{Becker2012}.
This is inconsistent with the $L_{\rm t}$ we observed.
On the other hand, the gas shock is expected to disappear at a very low accretion rate $L \ll L_{\rm coul}$, which might lead to a transition as well.
However, since it is still unknown about the physics in the shock, there are no theoretical calculations available to compare with observations.

While X-rays propagate in the highly magnetized plasma, both plasma and vacuum polarization effects are important for the cross section and therefore the radiation transfer in the neutron star atmosphere \citep[for a review, see][]{Harding2006}.
\cite{Wang1988} suggested that the relative importance of plasma and vacuum effects can be characterised by the ratio $\frac{\omega}{\delta_{\rm V}} = \frac{45\pi}{\alpha}\left( \frac{\omega_{\rm pe}}{\omega} \right)^2 \left( \frac{m_{\rm e}c^2}{E_{\rm cyc}} \right)^2$
, where $\omega_{\rm pe}=\sqrt{4\pi n_{\rm e}e^2/m_{\rm e}}$ is the plasma frequency, $m_{\rm e}$ is the election mass, $\omega$ is the frequency of the photons propagating in the medium, $\alpha$ is the fine structure constant, and $E_{\rm cyc}$ is the cyclotron line energy. 
If $\frac{\omega}{\delta_{\rm V}} \ll 1$, the X-ray propagation is mainly affected by the vacuum polarization, while the plasma effect dominates when $\frac{\omega}{\delta_{\rm V}} \gg 1$.
Both effects need to be taken into account for $\frac{\omega}{\delta_{\rm V}} \approx 1$.
For the transitional luminosity $L_{\rm t}$, the number density of electrons $n_{\rm e}$ can be estimated as $\frac{\dot{M}}{Svm_{\rm p}}$, where $m_{\rm p}$ is the proton mass and $S=5.2 \times 10^8$\,${\rm cm^2}$ is the area of spots at the neutron star surface calculated by using Equation 1 in \citet{Mushtukov2021}.
$v$ should be approximately the free-fall velocity $v \approx v_{\rm ff}=(\frac{2GM}{R})^{1/2}$, or $v \approx  v_{\rm ff}/4$ if a gas shock appears \citep{Bykov2004}.
For 1\,keV X-rays, $\frac{\omega}{\delta_{\rm V}}$ equals 2 and 0.5 in the cases with and without the gas shock, not significantly larger or smaller than unity.
This means that the transition around $L_{\rm t}$ might be related to the combination of plasma and vacuum effects, i.e., the former and the latter dominate the luminosity range $L \gtrsim L_{\rm t}$ and $L \lesssim L_{\rm t}$, respectively.
This possibility can be verified via polarization observations in the further, because the polarization angle is expected to be different between high and low states \citep{Lai2002}.

Another scenario for the transition might be related to interactions between the magnetosphere and outside accretion flows.
For example, it is known that when the accretion rate is very low, the system will evolve into the ``propeller regime" (i.e., the accreting matter is stopped by the centrifugal barrier by the dipole magnetic field), associated with significant temporal and spectral variations \citep{Campana2002, Tsygankov2016}.
However, the slow rotation of 1A 0535+262 makes the propeller effect only work when $L < L_{\rm prop} \approx 4\times10^{37}\,k^{7/2}\,B_{12}^2\,P^{-7/3}$\,$\rm erg\,s^{-1}$ $\approx 2\times 10^{33}$\,$\rm erg\,s^{-1}$, where $k$ is assumed to be $\sim$ 0.5 for a disk geometry \citep{Ghosh1978}, which is inconsistent with the transitional luminosity we detected.\\

The data are obtained from the High Energy Astrophysics Science Archive Research Center (HEASARC), provided by NASA's Goddard Space Flight Center.
This work is supported by the National Natural Science Foundation of China under grants No. 12173103.

\appendix

\renewcommand\thetable{\thesection.\arabic{table}}    
\setcounter{table}{0}  
\section{Spectral fitting parameters}

\begin{longrotatetable}
\begin{deluxetable*}{ccccccccccccc}
\tablecaption{Best-fitting spectral parameters of \textit{NICER} observations.}\label{data}
\tablewidth{700pt}
\renewcommand\arraystretch{1.4}
\tabletypesize{\scriptsize}
\tablehead{
\colhead{ObsID} & \colhead{Time} & 
\colhead{Exposure} & \colhead{$N_{\rm H}$} & 
\colhead{Photon Index} & \colhead{$Norm_{\rm Po}$} & 
\colhead{$T_{\rm BB}$} & \colhead{$Norm_{\rm BB}$} & 
\colhead{${E_{\rm line}}^*$} & \colhead{${\sigma_{\rm line}}^*$} & 
\colhead{Luminosity} & \colhead{$\chi^2$} & \colhead{dof} \\ 
\colhead{} & \colhead{(MJD)} & \colhead{(ks)} & \colhead{($10^{22}$ cm$^{-2}$)} & 
\colhead{($\Gamma$)} & \colhead{($10^{-2}$)} & \colhead{(keV)} &
\colhead{} & \colhead{(keV)} & \colhead{(keV)} & \colhead{($10^{35}\,$erg\,s$^{-1}$)} & \colhead{} & \colhead{}
} 
\startdata
1200360102 & 58386.28 & 1.469 & 0.66$^{+0.03}_{-0.03}$ & 0.69$^{+0.08}_{-0.05}$ & 2.15$^{+0.27}_{-0.23}$ & 1.25$^{+0.06}_{-0.04}$ & 10.6$^{+1.49}_{-1.19}$ & 6.29$^{+0.10}_{-0.00}$ & 0.183$^{+0.052}_{-0.052}$ & 10.84$^{+0.002}_{-0.109}$ & 112.82 & 126 \\ 
1200360103 & 58387.06 & 3.830 & 0.65$^{+0.01}_{-0.02}$ & 0.75$^{+0.03}_{-0.04}$ & 2.84$^{+0.15}_{-0.19}$ & 1.30$^{+0.04}_{-0.04}$ & 7.70$^{+0.99}_{-0.72}$ & 6.41$^{+0.07}_{-0.09}$ & 0.144$^{+0.097}_{-0.067}$ & 11.64$^{+0.002}_{-0.005}$ & 98.45 & 132 \\ 
1200360104 & 58388.93 & 0.986 & 0.64$^{+0.05}_{-0.05}$ & 0.73$^{+0.09}_{-0.09}$ & 2.23$^{+0.33}_{-0.35}$ & 1.24$^{+0.14}_{-0.09}$ & 6.33$^{+2.45}_{-1.91}$ & 6.40$^{+0.05}_{-0.07}$ & 0.033$^{+0.133}_{-0.028}$ & 9.085$^{+0.009}_{-0.007}$ & 93.59 & 118 \\ 
1200360105 & 58391.05 & 1.429 & 0.63$^{+0.04}_{-0.04}$ & 0.81$^{+0.10}_{-0.09}$ & 1.53$^{+0.19}_{-0.21}$ & 1.31$^{+0.11}_{-0.09}$ & 4.30$^{+1.10}_{-0.84}$ & 6.35$^{+0.10}_{-0.03}$ & 0.137$^{+0.105}_{-0.119}$ & 5.986$^{+0.008}_{-0.005}$ & 79.29 & 118 \\ 
2200360101 & 58724.38 & 1.029 & 0.67$^{+0.02}_{-0.01}$ & 0.86$^{+0.03}_{-0.03}$ & 13.7$^{+1.06}_{-0.58}$ & 1.20$^{+0.12}_{-0.13}$ & 11.6$^{+4.17}_{-4.43}$ & 6.36$^{+0.08}_{-0.04}$ & 0.137$^{+0.076}_{-0.101}$ & 37.92$^{+0.308}_{-0.159}$ & 93.23 & 126 \\ 
2200360102 & 58726.18 & 1.203 & 0.63$^{+0.01}_{-0.02}$ & 0.76$^{+0.03}_{-0.03}$ & 13.1$^{+0.57}_{-0.76}$ & 1.37$^{+0.15}_{-0.12}$ & 11.3$^{+3.60}_{-2.52}$ & 6.39$^{+0.06}_{-0.03}$ & 0.049$^{+0.144}_{-0.036}$ & 44.07$^{+0.147}_{-0.203}$ & 107.85 & 130 \\ 
2200360103 & 58728.18 & 1.422 & 0.63$^{+0.01}_{-0.01}$ & 0.79$^{+0.02}_{-0.03}$ & 16.6$^{+0.51}_{-0.59}$ & 1.68$^{+0.10}_{-0.13}$ & 8.30$^{+1.74}_{-1.16}$ & 6.39$^{+0.03}_{-0.02}$ & 0.001$^{+0.074}_{-0.001}$ & 55.62$^{+0.196}_{-0.228}$ & 96.20 & 134 \\ 
2200360104 & 58730.38 & 1.139 & 0.63$^{+0.01}_{-0.01}$ & 0.79$^{+0.04}_{-0.03}$ & 9.73$^{+0.46}_{-0.45}$ & 1.30$^{+0.16}_{-0.11}$ & 8.78$^{+2.33}_{-2.21}$ & 6.38$^{+0.05}_{-0.05}$ & 0.073$^{+0.157}_{-0.062}$ & 30.66$^{+0.180}_{-0.171}$ & 93.75 & 126 \\ 
2200360105 & 58732.51 & 1.020 & 0.61$^{+0.01}_{-0.01}$ & 0.81$^{+0.05}_{-0.05}$ & 6.97$^{+0.35}_{-0.51}$ & 1.26$^{+0.22}_{-0.13}$ & 8.30$^{+4.15}_{-2.49}$ & 6.39$^{+0.08}_{-0.08}$ & 0.250$^{+0.003}_{-0.182}$ & 22.12$^{+0.127}_{-0.180}$ & 109.43 & 124 \\ 
2200360107 & 58739.01 & 1.474 & 0.65$^{+0.01}_{-0.02}$ & 1.05$^{+0.04}_{-0.06}$ & 0.84$^{+0.05}_{-0.06}$ & 1.25$^{+0.10}_{-0.11}$ & 1.77$^{+0.54}_{-0.41}$ & 6.50$^{+0.00}_{-0.19}$ & 0.074$^{+0.026}_{-0.019}$ & 2.159$^{+0.003}_{-0.002}$ & 95.98 & 107 \\ 
2200360108 & 58742.16 & 3.726 & 0.49$^{+0.00}_{-0.03}$ & 0.54$^{+0.00}_{-0.04}$ & 0.48$^{+0.00}_{-0.04}$ & 0.99$^{+0.01}_{-0.00}$ & 4.09$^{+0.07}_{-0.00}$ & 6.29$^{+0.18}_{-0.00}$ & 0.250$^{+0.000}_{-0.028}$ & 2.634$^{+0.003}_{-0.000}$ & 109.73 & 120 \\ 
2200360109 & 58743.21 & 1.315 & 0.60$^{+0.06}_{-0.05}$ & 0.98$^{+0.15}_{-0.10}$ & 0.57$^{+0.10}_{-0.08}$ & 1.13$^{+0.11}_{-0.06}$ & 2.24$^{+0.54}_{-0.54}$ & 6.42$^{+0.04}_{-0.09}$ & 0.001$^{+0.223}_{-0.003}$ & 1.711$^{+0.003}_{-0.004}$ & 66.85&  96 \\ 
2200360110 & 58743.97 & 2.011 & 0.66$^{+0.06}_{-0.02}$ & 1.04$^{+0.08}_{-0.00}$ & 0.77$^{+0.11}_{-0.07}$ & 1.22$^{+0.07}_{-0.00}$ & 2.10$^{+0.29}_{-0.44}$ & 6.36$^{+0.12}_{-0.03}$ & 0.250$^{+0.005}_{-0.186}$ & 2.132$^{+0.002}_{-0.003}$ & 111.82 & 110 \\ 
2200360111 & 58746.05 & 1.554 & 0.59$^{+0.03}_{-0.06}$ & 1.01$^{+0.10}_{-0.11}$ & 0.47$^{+0.06}_{-0.09}$ & 1.09$^{+0.09}_{-0.08}$ & 1.86$^{+0.45}_{-0.37}$ & 6.29$^{+0.18}_{-0.00}$ & 0.156$^{+0.085}_{-0.086}$ & 1.323$^{+0.001}_{-0.003}$ & 76.87&  96 \\ 
2200360112 & 58749.00 & 2.394 & 0.67$^{+0.06}_{-0.06}$ & 1.17$^{+0.16}_{-0.10}$ & 0.57$^{+0.08}_{-0.09}$ & 1.17$^{+0.06}_{-0.05}$ & 1.70$^{+0.43}_{-0.20}$ & 6.37$^{+0.11}_{-0.06}$ & 0.060$^{+0.172}_{-0.051}$ & 1.326$^{+0.001}_{-0.003}$ & 83.64 & 107 \\ 
2200360113 & 58751.07 & 2.331 & 0.61$^{+0.06}_{-0.03}$ & 0.95$^{+0.08}_{-0.05}$ & 0.61$^{+0.11}_{-0.03}$ & 1.11$^{+0.05}_{-0.08}$ & 2.40$^{+0.43}_{-0.54}$ & 6.50$^{+0.00}_{-0.12}$ & 0.250$^{+0.003}_{-0.166}$ & 1.884$^{+0.002}_{-0.001}$ & 124.12 & 112 \\ 
2200360114 & 58753.91 & 2.176 & 0.51$^{+0.01}_{-0.03}$ & 0.63$^{+0.00}_{-0.07}$ & 0.46$^{+0.00}_{-0.06}$ & 0.90$^{+0.00}_{-0.02}$ & 3.07$^{+0.76}_{-0.01}$ & 6.49$^{+0.00}_{-0.19}$ & 0.249$^{+0.000}_{-0.025}$ & 2.026$^{+0.001}_{-0.000}$ & 129.18 & 115 \\ 
2200360115 & 58755.91 & 1.932 & 0.61$^{+0.01}_{-0.03}$ & 0.94$^{+0.05}_{-0.07}$ & 0.72$^{+0.05}_{-0.08}$ & 1.19$^{+0.08}_{-0.07}$ & 2.17$^{+0.36}_{-0.40}$ & 6.50$^{+0.00}_{-0.16}$ & 0.241$^{+0.001}_{-0.205}$ & 2.228$^{+0.001}_{-0.001}$ & 121.65 & 113 \\ 
2200360116 & 58758.81 & 1.755 & 0.62$^{+0.00}_{-0.03}$ & 1.09$^{+0.02}_{-0.01}$ & 0.48$^{+0.03}_{-0.03}$ & 1.18$^{+0.02}_{-0.00}$ & 1.38$^{+0.00}_{-0.06}$ & 6.43$^{+0.01}_{-0.07}$ & 0.001$^{+0.116}_{-0.000}$ & 1.230$^{+0.000}_{-0.002}$ & 110.14 & 100 \\
2200360118 & 58762.44 & 0.969 & 0.65$^{+0.11}_{-0.13}$ & 1.06$^{+0.15}_{-0.24}$ & 0.35$^{+0.10}_{-0.12}$ & 0.99$^{+0.11}_{-0.11}$ & 1.48$^{+0.80}_{-0.66}$ & 6.29$^{+0.18}_{-0.00}$ & 0.112$^{+0.118}_{-0.080}$ & 0.855$^{+0.003}_{-0.001}$ & 86.31&  89 \\ 
3200360101 & 58934.44 & 6.926 & 0.73$^{+0.03}_{-0.05}$ & 0.96$^{+0.08}_{-0.08}$ & 1.03$^{+0.11}_{-0.14}$ & 1.27$^{+0.07}_{-0.07}$ & 3.04$^{+0.71}_{-0.46}$ & 6.43$^{+0.06}_{-0.04}$ & 0.036$^{+0.200}_{-0.001}$ & 3.288$^{+0.002}_{-0.001}$ & 142.18 & 124 \\ 
3200360102 & 58935.02 & 12.56 & 0.66$^{+0.02}_{-0.01}$ & 0.86$^{+0.05}_{-0.03}$ & 1.24$^{+0.10}_{-0.06}$ & 1.26$^{+0.05}_{-0.03}$ & 3.99$^{+0.30}_{-0.49}$ & 6.42$^{+0.06}_{-0.07}$ & 0.112$^{+0.123}_{-0.049}$ & 4.527$^{+0.001}_{-0.001}$ & 98.84 & 133 \\ 
3200360103 & 58935.99 & 3.669 & 0.68$^{+0.02}_{-0.03}$ & 0.96$^{+0.06}_{-0.06}$ & 1.23$^{+0.08}_{-0.11}$ & 1.26$^{+0.05}_{-0.06}$ & 3.36$^{+0.62}_{-0.36}$ & 6.42$^{+0.06}_{-0.08}$ & 0.106$^{+0.134}_{-0.052}$ & 3.821$^{+0.002}_{-0.002}$ & 105.54 & 120 \\ 
3200360104 & 58937.15 & 4.138 & 0.68$^{+0.04}_{-0.03}$ & 0.89$^{+0.09}_{-0.05}$ & 1.03$^{+0.15}_{-0.10}$ & 1.31$^{+0.07}_{-0.05}$ & 3.22$^{+0.48}_{-0.53}$ & 6.35$^{+0.10}_{-0.04}$ & 0.051$^{+0.179}_{-0.035}$ & 3.780$^{+0.001}_{-0.004}$ & 121.60 & 121 \\ 
3200360105 & 58940.25 & 4.456 & 0.68$^{+0.00}_{-0.02}$ & 0.83$^{+0.01}_{-0.03}$ & 0.97$^{+0.00}_{-0.04}$ & 1.24$^{+0.02}_{-0.00}$ & 3.94$^{+0.00}_{-0.09}$ & 6.50$^{+0.00}_{-0.10}$ & 0.250$^{+0.000}_{-0.000}$ & 3.876$^{+0.003}_{-0.002}$ & 100.15 & 122 \\ 
3200360106 & 58941.35 & 4.941 & 0.69$^{+0.03}_{-0.03}$ & 0.82$^{+0.07}_{-0.09}$ & 0.96$^{+0.09}_{-0.13}$ & 1.21$^{+0.06}_{-0.07}$ & 3.88$^{+0.97}_{-0.48}$ & 6.43$^{+0.03}_{-0.07}$ & 0.011$^{+0.194}_{-0.006}$ & 3.771$^{+0.002}_{-0.002}$ & 99.81 & 121 \\ 
3200360107 & 58942.38 & 4.206 & 0.68$^{+0.04}_{-0.02}$ & 0.61$^{+0.09}_{-0.05}$ & 1.42$^{+0.24}_{-0.12}$ & 1.24$^{+0.05}_{-0.03}$ & 9.02$^{+0.84}_{-1.41}$ & 6.37$^{+0.05}_{-0.04}$ & 0.086$^{+0.082}_{-0.048}$ & 8.484$^{+0.005}_{-0.004}$ & 123.62 & 124 \\ 
3200360108 & 58943.35 & 4.120 & 0.69$^{+0.01}_{-0.01}$ & 0.65$^{+0.03}_{-0.04}$ & 3.92$^{+0.25}_{-0.34}$ & 1.22$^{+0.02}_{-0.04}$ & 15.8$^{+2.41}_{-1.38}$ & 6.39$^{+0.04}_{-0.03}$ & 0.076$^{+0.124}_{-0.036}$ & 18.93$^{+0.005}_{-0.008}$ & 118.42 & 133 \\ 
3200360109 & 58944.84 & 1.901 & 0.64$^{+0.01}_{-0.02}$ & 0.63$^{+0.03}_{-0.05}$ & 4.89$^{+0.36}_{-0.41}$ & 1.21$^{+0.04}_{-0.05}$ & 15.9$^{+2.89}_{-2.36}$ & 6.36$^{+0.11}_{-0.05}$ & 0.250$^{+0.001}_{-0.151}$ & 22.84$^{+0.163}_{-0.008}$ & 120.68 & 126 \\ 
3200360110 & 58945.42 & 1.532 & 0.70$^{+0.01}_{-0.02}$ & 0.73$^{+0.02}_{-0.04}$ & 6.67$^{+0.41}_{-0.44}$ & 1.22$^{+0.03}_{-0.07}$ & 13.1$^{+2.93}_{-1.86}$ & 6.29$^{+0.11}_{-0.00}$ & 0.250$^{+0.006}_{-0.154}$ & 24.86$^{+0.155}_{-0.009}$ & 135.78 & 124 \\ 
3200360111 & 58946.39 & 2.588 & 0.62$^{+0.01}_{-0.01}$ & 0.77$^{+0.02}_{-0.03}$ & 9.03$^{+0.44}_{-0.51}$ & 1.13$^{+0.04}_{-0.05}$ & 16.9$^{+3.43}_{-2.34}$ & 6.39$^{+0.05}_{-0.03}$ & 0.061$^{+0.178}_{-0.029}$ & 30.23$^{+0.115}_{-0.118}$ & 110.75 & 134 \\ 
3200360112 & 58948.20 & 4.072 & 0.65$^{+0.01}_{-0.01}$ & 0.86$^{+0.03}_{-0.03}$ & 8.71$^{+0.31}_{-0.46}$ & 1.37$^{+0.13}_{-0.10}$ & 7.02$^{+2.24}_{-1.44}$ & 6.42$^{+0.03}_{-0.03}$ & 0.045$^{+0.154}_{-0.035}$ & 25.06$^{+0.006}_{-0.008}$ & 88.15 & 135 \\ 
3200360113 & 58948.97 & 4.121 & 0.63$^{+0.02}_{-0.01}$ & 0.81$^{+0.03}_{-0.01}$ & 8.60$^{+0.52}_{-0.28}$ & 1.23$^{+0.09}_{-0.05}$ & 10.7$^{+1.44}_{-2.41}$ & 6.37$^{+0.07}_{-0.01}$ & 0.021$^{+0.189}_{-0.016}$ & 26.99$^{+0.008}_{-0.136}$ & 95.79 & 132 \\ 
3200360156 & 59206.92 & 0.949 & 0.67$^{+0.01}_{-0.01}$ & 1.09$^{+0.07}_{-0.04}$ & 13.9$^{+0.40}_{-0.50}$ & 1.67$^{+0.16}_{-0.14}$ & 5.19$^{+1.15}_{-0.50}$ & 6.38$^{+0.09}_{-0.07}$ & 0.045$^{+0.180}_{-0.032}$ & 29.10$^{+0.131}_{-0.263}$ & 110.14 & 130 \\ 
3200360157 & 59206.99 & 1.644 & 0.68$^{+0.00}_{-0.01}$ & 1.09$^{+0.02}_{-0.04}$ & 11.7$^{+0.25}_{-0.45}$ & 1.56$^{+0.07}_{-0.13}$ & 6.38$^{+1.07}_{-0.86}$ & 6.48$^{+0.01}_{-0.06}$ & 0.075$^{+0.145}_{-0.065}$ & 25.12$^{+0.007}_{-0.140}$ & 112.17 & 131 \\ 
3200360158 & 59208.80 & 0.901 & 0.66$^{+0.02}_{-0.02}$ & 0.97$^{+0.06}_{-0.04}$ & 6.28$^{+0.50}_{-0.35}$ & 1.29$^{+0.12}_{-0.08}$ & 9.43$^{+1.93}_{-2.14}$ & 6.40$^{+0.07}_{-0.09}$ & 0.212$^{+0.032}_{-0.147}$ & 16.85$^{+0.121}_{-0.156}$ & 117.17 & 125 \\ 
3200360159 & 59209.17 & 1.954 & 0.67$^{+0.01}_{-0.02}$ & 0.99$^{+0.04}_{-0.05}$ & 4.50$^{+0.21}_{-0.34}$ & 1.26$^{+0.06}_{-0.07}$ & 7.23$^{+1.68}_{-0.83}$ & 6.42$^{+0.06}_{-0.09}$ & 0.084$^{+0.160}_{-0.040}$ & 11.55$^{+0.004}_{-0.009}$ & 106.53 & 127 \\ 
3200360160 & 59210.85 & 1.584 & 0.63$^{+0.01}_{-0.02}$ & 0.98$^{+0.03}_{-0.05}$ & 2.30$^{+0.14}_{-0.16}$ & 1.26$^{+0.04}_{-0.06}$ & 5.02$^{+0.72}_{-0.56}$ & 6.40$^{+0.07}_{-0.09}$ & 0.249$^{+0.005}_{-0.191}$ & 6.606$^{+0.005}_{-0.003}$ & 109.21 & 121 \\ 
3200360161 & 59213.31 & 0.306 & 0.66$^{+0.07}_{-0.06}$ & 1.06$^{+0.06}_{-0.04}$ & 1.26$^{+0.23}_{-0.20}$ & 1.08$^{+0.05}_{-0.08}$ & 4.86$^{+0.82}_{-0.27}$ & 6.29$^{+0.18}_{-0.00}$ & 0.250$^{+0.000}_{-0.000}$ & 3.265$^{+0.008}_{-0.121}$ & 81.76&  82 \\ 
3200360162 & 59214.27 & 7.279 & 0.64$^{+0.02}_{-0.02}$ & 1.01$^{+0.05}_{-0.04}$ & 1.19$^{+0.08}_{-0.10}$ & 1.13$^{+0.05}_{-0.04}$ & 3.57$^{+0.67}_{-0.45}$ & 6.34$^{+0.10}_{-0.03}$ & 0.001$^{+0.226}_{-0.008}$ & 3.154$^{+0.001}_{-0.002}$ & 96.15 & 129 \\ 
3200360163 & 59214.98 & 3.633 & 0.63$^{+0.04}_{-0.01}$ & 1.06$^{+0.08}_{-0.02}$ & 0.92$^{+0.11}_{-0.03}$ & 1.08$^{+0.04}_{-0.02}$ & 3.39$^{+0.32}_{-0.53}$ & 6.49$^{+0.00}_{-0.17}$ & 0.250$^{+0.000}_{-0.056}$ & 2.330$^{+0.000}_{-0.003}$ & 100.51 & 119 \\ 
3200360164 & 59216.33 & 12.34 & 0.68$^{+0.03}_{-0.01}$ & 1.12$^{+0.05}_{-0.04}$ & 1.10$^{+0.09}_{-0.06}$ & 1.14$^{+0.05}_{-0.03}$ & 3.06$^{+0.34}_{-0.47}$ & 6.32$^{+0.13}_{-0.01}$ & 0.250$^{+0.003}_{-0.185}$ & 2.521$^{+0.000}_{-0.001}$ & 93.29 & 130 \\ 
3200360165 & 59217.36 & 6.985 & 0.68$^{+0.04}_{-0.03}$ & 1.11$^{+0.07}_{-0.06}$ & 1.24$^{+0.14}_{-0.10}$ & 1.19$^{+0.07}_{-0.04}$ & 3.48$^{+0.46}_{-0.59}$ & 6.36$^{+0.09}_{-0.04}$ & 0.250$^{+0.003}_{-0.165}$ & 3.110$^{+0.001}_{-0.001}$ & 118.57 & 128 \\ 
3200360166 & 59218.20 & 3.524 & 0.67$^{+0.03}_{-0.03}$ & 1.09$^{+0.05}_{-0.06}$ & 0.97$^{+0.07}_{-0.11}$ & 1.06$^{+0.06}_{-0.04}$ & 3.49$^{+0.67}_{-0.62}$ & 6.29$^{+0.08}_{-0.00}$ & 0.250$^{+0.000}_{-0.045}$ & 2.280$^{+0.001}_{-0.000}$ & 92.87 & 119 \\ 
3200360167 & 59218.99 & 6.149 & 0.73$^{+0.03}_{-0.03}$ & 1.22$^{+0.06}_{-0.06}$ & 1.17$^{+0.10}_{-0.09}$ & 1.23$^{+0.06}_{-0.05}$ & 2.44$^{+0.37}_{-0.37}$ & 6.35$^{+0.09}_{-0.04}$ & 0.066$^{+0.152}_{-0.042}$ & 2.424$^{+0.001}_{-0.001}$ & 128.41 & 124 \\ 
3200360168 & 59220.07 & 1.881 & 0.62$^{+0.01}_{-0.00}$ & 1.11$^{+0.01}_{-0.06}$ & 0.72$^{+0.00}_{-0.00}$ & 1.10$^{+0.00}_{-0.10}$ & 2.41$^{+0.51}_{-0.01}$ & 6.50$^{+0.00}_{-0.07}$ & 0.204$^{+0.001}_{-0.060}$ & 1.737$^{+0.001}_{-0.000}$ & 95.95 & 107 \\ 
3200360169 & 59221.04 & 5.710 & 0.64$^{+0.03}_{-0.03}$ & 1.09$^{+0.04}_{-0.06}$ & 0.70$^{+0.06}_{-0.06}$ & 1.02$^{+0.00}_{-0.03}$ & 2.55$^{+0.19}_{-0.12}$ & 6.34$^{+0.03}_{-0.04}$ & 0.001$^{+0.060}_{-0.000}$ & 1.602$^{+0.001}_{-0.000}$ & 104.53 & 119 \\ 
3200360170 & 59222.20 & 2.535 & 0.65$^{+0.04}_{-0.03}$ & 1.08$^{+0.07}_{-0.06}$ & 0.86$^{+0.10}_{-0.09}$ & 1.12$^{+0.08}_{-0.06}$ & 2.99$^{+0.51}_{-0.67}$ & 6.34$^{+0.13}_{-0.03}$ & 0.104$^{+0.134}_{-0.086}$ & 2.198$^{+0.002}_{-0.001}$ & 128.08 & 113 \\ 
3200360171 & 59223.04 & 3.150 & 0.69$^{+0.05}_{-0.03}$ & 1.24$^{+0.10}_{-0.06}$ & 0.84$^{+0.12}_{-0.07}$ & 1.18$^{+0.09}_{-0.04}$ & 2.22$^{+0.21}_{-0.55}$ & 6.42$^{+0.06}_{-0.10}$ & 0.250$^{+0.004}_{-0.185}$ & 1.779$^{+0.002}_{-0.001}$ & 109.34 & 114 \\ 
3200360172 & 59224.01 & 2.135 & 0.75$^{+0.08}_{-0.01}$ & 1.40$^{+0.31}_{-0.04}$ & 0.79$^{+0.18}_{-0.04}$ & 1.17$^{+0.13}_{-0.03}$ & 2.17$^{+0.28}_{-0.35}$ & 6.31$^{+0.17}_{-0.01}$ & 0.250$^{+0.002}_{-0.101}$ & 1.424$^{+0.000}_{-0.004}$ & 96.25 & 104 \\ 
3200360173 & 59225.04 & 3.253 & 0.57$^{+0.04}_{-0.04}$ & 0.85$^{+0.07}_{-0.06}$ & 0.63$^{+0.09}_{-0.07}$ & 0.97$^{+0.05}_{-0.03}$ & 2.95$^{+0.52}_{-0.72}$ & 6.41$^{+0.06}_{-0.08}$ & 0.001$^{+0.224}_{-0.003}$ & 1.973$^{+0.001}_{-0.001}$ & 111.73 & 116 \\ 
3200360174 & 59226.21 & 2.255 & 0.63$^{+0.01}_{-0.01}$ & 0.94$^{+0.01}_{-0.01}$ & 0.90$^{+0.00}_{-0.00}$ & 1.00$^{+0.00}_{-0.00}$ & 4.70$^{+0.08}_{-0.00}$ & 6.29$^{+0.19}_{-0.00}$ & 0.250$^{+0.000}_{-0.000}$ & 2.649$^{+0.002}_{-0.003}$ & 132.01  & 116 \\ 
3200360175 & 59227.05 & 3.525 & 0.66$^{+0.04}_{-0.01}$ & 1.15$^{+0.12}_{-0.00}$ & 0.64$^{+0.07}_{-0.02}$ & 1.08$^{+0.08}_{-0.04}$ & 2.02$^{+0.00}_{-0.11}$ & 6.50$^{+0.00}_{-0.02}$ & 0.250$^{+0.001}_{-0.031}$ & 1.404$^{+0.000}_{-0.003}$ & 92.31 & 111 \\ 
3200360176 & 59228.73 & 0.879 & 0.67$^{+0.07}_{-0.07}$ & 1.13$^{+0.22}_{-0.20}$ & 0.76$^{+0.14}_{-0.18}$ & 1.09$^{+0.15}_{-0.13}$ & 2.61$^{+1.46}_{-0.59}$ & 6.41$^{+0.05}_{-0.10}$ & 0.023$^{+0.220}_{-0.006}$ & 1.761$^{+0.004}_{-0.004}$ & 92.04&  90 \\ 
3200360177 & 59229.31 & 4.739 & 0.70$^{+0.04}_{-0.03}$ & 1.18$^{+0.08}_{-0.05}$ & 0.69$^{+0.08}_{-0.06}$ & 1.05$^{+0.09}_{-0.04}$ & 1.94$^{+0.39}_{-0.44}$ & 6.45$^{+0.04}_{-0.03}$ & 0.001$^{+0.188}_{-0.004}$ & 1.359$^{+0.000}_{-0.001}$ & 92.73 & 116 \\ 
3200360178 & 59230.73 & 1.373 & 0.69$^{+0.01}_{-0.05}$ & 1.25$^{+0.02}_{-0.14}$ & 0.76$^{+0.04}_{-0.11}$ & 1.18$^{+0.05}_{-0.09}$ & 1.93$^{+0.45}_{-0.36}$ & 6.50$^{+0.00}_{-0.03}$ & 0.165$^{+0.062}_{-0.124}$ & 1.553$^{+0.003}_{-0.001}$ & 109.93&  97 \\ 
3200360179 & 59231.31 & 5.363 & 0.68$^{+0.05}_{-0.02}$ & 1.30$^{+0.08}_{-0.09}$ & 0.57$^{+0.06}_{-0.04}$ & 1.12$^{+0.03}_{-0.07}$ & 1.64$^{+0.23}_{-0.17}$ & 6.38$^{+0.09}_{-0.00}$ & 0.007$^{+0.076}_{-0.005}$ & 1.085$^{+0.001}_{-0.001}$ & 110.90 & 116 \\ 
3200360180 & 59232.54 & 2.466 & 0.71$^{+0.00}_{-0.03}$ & 1.32$^{+0.00}_{-0.10}$ & 0.74$^{+0.00}_{-0.08}$ & 1.17$^{+0.00}_{-0.08}$ & 1.85$^{+0.38}_{-0.00}$ & 6.29$^{+0.03}_{-0.00}$ & 0.250$^{+0.000}_{-0.044}$ & 1.401$^{+0.000}_{-0.000}$ & 125.70 & 108 \\ 
3200360181 & 59233.05 & 4.071 & 0.72$^{+0.05}_{-0.05}$ & 1.29$^{+0.11}_{-0.04}$ & 0.66$^{+0.07}_{-0.07}$ & 1.15$^{+0.07}_{-0.01}$ & 1.67$^{+0.15}_{-0.21}$ & 6.50$^{+0.00}_{-0.10}$ & 0.250$^{+0.002}_{-0.133}$ & 1.255$^{+0.000}_{-0.002}$ & 115.67 & 114 \\ 
3200360182 & 59234.28 & 2.699 & 0.61$^{+0.05}_{-0.04}$ & 1.10$^{+0.13}_{-0.11}$ & 0.52$^{+0.07}_{-0.07}$ & 1.07$^{+0.10}_{-0.05}$ & 2.08$^{+0.49}_{-0.41}$ & 6.39$^{+0.09}_{-0.08}$ & 0.250$^{+0.006}_{-0.202}$ & 1.281$^{+0.001}_{-0.001}$ & 112.42 & 109 \\ 
3200360183 & 59234.99 & 8.205 & 0.76$^{+0.03}_{-0.06}$ & 1.41$^{+0.03}_{-0.13}$ & 0.66$^{+0.06}_{-0.08}$ & 1.20$^{+0.06}_{-0.07}$ & 1.27$^{+0.23}_{-0.30}$ & 6.50$^{+0.00}_{-0.15}$ & 0.250$^{+0.004}_{-0.172}$ & 1.081$^{+0.001}_{-0.000}$ & 115.67 & 120 \\ 
3200360184 & 59236.02 & 4.112 & 0.65$^{+0.04}_{-0.05}$ & 1.21$^{+0.10}_{-0.11}$ & 0.46$^{+0.05}_{-0.07}$ & 1.04$^{+0.05}_{-0.06}$ & 1.80$^{+0.40}_{-0.20}$ & 6.38$^{+0.06}_{-0.08}$ & 0.054$^{+0.186}_{-0.041}$ & 0.966$^{+0.001}_{-0.001}$ & 81.31 & 113 \\ 
3200360185 & 59237.38 & 2.576 & 0.70$^{+0.05}_{-0.00}$ & 1.36$^{+0.16}_{-0.00}$ & 0.47$^{+0.05}_{-0.00}$ & 1.15$^{+0.08}_{-0.00}$ & 1.21$^{+0.08}_{-0.21}$ & 6.50$^{+0.00}_{-0.02}$ & 0.250$^{+0.000}_{-0.004}$ & 0.844$^{+0.000}_{-0.002}$ & 107.25 & 100 \\ 
3200360186 & 59238.09 & 3.822 & 0.62$^{+0.07}_{-0.04}$ & 1.01$^{+0.12}_{-0.11}$ & 0.44$^{+0.09}_{-0.07}$ & 0.90$^{+0.09}_{-0.04}$ & 2.46$^{+0.76}_{-0.79}$ & 6.50$^{+0.00}_{-0.13}$ & 0.055$^{+0.156}_{-0.038}$ & 1.085$^{+0.001}_{-0.002}$ & 99.48 & 114 \\ 
3200360187 & 59240.67 & 0.669 & 0.75$^{+0.01}_{-0.00}$ & 1.57$^{+0.02}_{-0.00}$ & 0.62$^{+0.01}_{-0.00}$ & 1.20$^{+0.03}_{-0.00}$ & 1.57$^{+0.00}_{-0.01}$ & 6.33$^{+0.15}_{-0.03}$ & 0.250$^{+0.000}_{-0.000}$ & 0.997$^{+0.003}_{-0.000}$ & 53.39&  77 \\ 
3200360188 & 59241.12 & 4.823 & 0.70$^{+0.04}_{-0.03}$ & 1.29$^{+0.05}_{-0.10}$ & 0.52$^{+0.03}_{-0.02}$ & 1.18$^{+0.07}_{-0.06}$ & 0.98$^{+0.02}_{-0.15}$ & 6.50$^{+0.00}_{-0.14}$ & 0.250$^{+0.003}_{-0.234}$ & 0.934$^{+0.002}_{-0.000}$ & 91.90 & 116 \\ 
3200360189 & 59242.28 & 3.415 & 0.67$^{+0.06}_{-0.04}$ & 1.24$^{+0.11}_{-0.10}$ & 0.47$^{+0.08}_{-0.05}$ & 1.07$^{+0.07}_{-0.06}$ & 1.61$^{+0.25}_{-0.38}$ & 6.30$^{+0.14}_{-0.00}$ & 0.123$^{+0.105}_{-0.087}$ & 0.950$^{+0.001}_{-0.001}$ & 100.68 & 111 \\ 
3200360190 & 59242.99 & 5.440 & 0.72$^{+0.04}_{-0.04}$ & 1.28$^{+0.02}_{-0.12}$ & 0.57$^{+0.06}_{-0.06}$ & 1.09$^{+0.05}_{-0.12}$ & 1.18$^{+0.35}_{-0.31}$ & 6.50$^{+0.00}_{-0.15}$ & 0.152$^{+0.060}_{-0.143}$ & 0.963$^{+0.002}_{-0.000}$ & 111.23 & 117 \\ 
3200360191 & 59246.22 & 1.536 & 0.65$^{+0.05}_{-0.04}$ & 1.19$^{+0.09}_{-0.09}$ & 0.55$^{+0.04}_{-0.03}$ & 1.14$^{+0.02}_{-0.03}$ & 1.59$^{+0.02}_{-0.02}$ & 6.29$^{+0.19}_{-0.00}$ & 0.250$^{+0.000}_{-0.000}$ & 1.198$^{+0.002}_{-0.001}$ & 85.14&  97 \\ 
3200360192 & 59248.03 & 0.975 & 0.81$^{+0.10}_{-0.11}$ & 1.82$^{+0.39}_{-0.36}$ & 1.02$^{+0.22}_{-0.21}$ & 1.22$^{+0.01}_{-0.04}$ & 3.25$^{+0.67}_{-0.55}$ & 6.50$^{+0.00}_{-0.16}$ & 0.243$^{+0.001}_{-0.190}$ & 1.629$^{+0.004}_{-0.004}$ & 84.93 & 109 \\ 
3200360193 & 59250.03 & 1.138 & 0.81$^{+0.03}_{-0.01}$ & 1.51$^{+0.00}_{-0.02}$ & 1.09$^{+0.02}_{-0.00}$ & 1.34$^{+0.02}_{-0.01}$ & 1.32$^{+0.00}_{-0.04}$ & 6.29$^{+0.19}_{-0.00}$ & 0.001$^{+0.000}_{-0.000}$ & 1.566$^{+0.003}_{-0.000}$ & 85.92 & 105 \\ 
3200360194 & 59251.96 & 1.537 & 0.56$^{+0.06}_{-0.04}$ & 0.69$^{+0.08}_{-0.07}$ & 0.85$^{+0.17}_{-0.11}$ & 0.94$^{+0.04}_{-0.03}$ & 6.86$^{+0.64}_{-1.14}$ & 6.40$^{+0.01}_{-0.05}$ & 0.001$^{+0.238}_{-0.007}$ & 3.643$^{+0.006}_{-0.005}$ & 86.34 & 118 \\ 
3200360195 & 59252.16 & 1.318 & 0.55$^{+0.00}_{-0.02}$ & 0.69$^{+0.01}_{-0.02}$ & 0.79$^{+0.00}_{-0.08}$ & 0.95$^{+0.07}_{-0.00}$ & 5.47$^{+0.00}_{-0.33}$ & 6.29$^{+0.04}_{-0.00}$ & 0.185$^{+0.008}_{-0.000}$ & 3.306$^{+0.000}_{-0.003}$ & 92.45 & 114 \\ 
3200360196 & 59253.58 & 6.414 & 0.68$^{+0.02}_{-0.02}$ & 0.97$^{+0.04}_{-0.04}$ & 2.95$^{+0.18}_{-0.19}$ & 1.23$^{+0.04}_{-0.03}$ & 8.31$^{+0.86}_{-0.78}$ & 6.38$^{+0.07}_{-0.03}$ & 0.037$^{+0.187}_{-0.026}$ & 8.812$^{+0.002}_{-0.002}$ & 79.51 & 136 \\ 
3200360197 & 59254.03 & 1.564 & 0.69$^{+0.03}_{-0.03}$ & 0.95$^{+0.07}_{-0.06}$ & 3.75$^{+0.37}_{-0.40}$ & 1.16$^{+0.06}_{-0.05}$ & 13.6$^{+2.45}_{-1.91}$ & 6.36$^{+0.08}_{-0.04}$ & 0.148$^{+0.090}_{-0.054}$ & 11.66$^{+0.100}_{-0.116}$ & 113.79 & 127 \\ 
3200360198 & 59256.03 & 1.565 & 0.72$^{+0.02}_{-0.02}$ & 0.99$^{+0.04}_{-0.04}$ & 4.37$^{+0.27}_{-0.34}$ & 1.25$^{+0.05}_{-0.06}$ & 10.0$^{+1.96}_{-1.07}$ & 6.39$^{+0.08}_{-0.07}$ & 0.002$^{+0.234}_{-0.024}$ & 12.27$^{+0.003}_{-0.009}$ & 77.95 & 127 \\
3200360199 & 59257.07 & 1.496 & 0.71$^{+0.03}_{-0.03}$ & 0.97$^{+0.07}_{-0.04}$ & 4.12$^{+0.35}_{-0.37}$ & 1.19$^{+0.10}_{-0.05}$ & 10.5$^{+2.41}_{-2.00}$ & 6.37$^{+0.09}_{-0.05}$ & 0.066$^{+0.164}_{-0.028}$ & 11.55$^{+0.005}_{-0.008}$ & 96.34 & 125 \\ 
3200360201 & 59259.71 & 1.757 & 0.70$^{+0.01}_{-0.03}$ & 0.96$^{+0.03}_{-0.05}$ & 5.10$^{+0.27}_{-0.43}$ & 1.15$^{+0.05}_{-0.05}$ & 11.0$^{+2.38}_{-1.61}$ & 6.34$^{+0.10}_{-0.02}$ & 0.083$^{+0.160}_{-0.057}$ & 13.47$^{+0.007}_{-0.009}$ & 98.21 & 127 \\ 
3200360202 & 59262.16 & 0.940 & 0.62$^{+0.03}_{-0.01}$ & 0.83$^{+0.09}_{-0.04}$ & 2.61$^{+0.35}_{-0.22}$ & 1.16$^{+0.10}_{-0.05}$ & 9.33$^{+1.91}_{-2.34}$ & 6.42$^{+0.03}_{-0.10}$ & 0.002$^{+0.230}_{-0.004}$ & 9.293$^{+0.005}_{-0.104}$ & 108.76 & 116 \\ 
3200360203 & 59263.00 & 1.603 & 0.66$^{+0.02}_{-0.02}$ & 1.00$^{+0.08}_{-0.03}$ & 2.13$^{+0.19}_{-0.17}$ & 1.21$^{+0.08}_{-0.01}$ & 6.33$^{+0.51}_{-0.91}$ & 6.34$^{+0.14}_{-0.03}$ & 0.250$^{+0.001}_{-0.116}$ & 6.165$^{+0.001}_{-0.007}$ & 106.67 & 117 \\ 
3200360204 & 59264.37 & 0.283 & 0.64$^{+0.04}_{-0.05}$ & 0.87$^{+0.10}_{-0.11}$ & 2.67$^{+0.33}_{-0.47}$ & 1.24$^{+0.11}_{-0.06}$ & 9.16$^{+2.00}_{-2.43}$ & 6.37$^{+0.08}_{-0.06}$ & 0.146$^{+0.080}_{-0.041}$ & 9.723$^{+0.222}_{-0.009}$ & 88.60&  99 \\ 
3200360205 & 59265.14 & 0.483 & 0.66$^{+0.06}_{-0.05}$ & 1.02$^{+0.16}_{-0.14}$ & 2.09$^{+0.38}_{-0.37}$ & 1.25$^{+0.12}_{-0.09}$ & 6.78$^{+1.99}_{-1.43}$ & 6.47$^{+0.01}_{-0.12}$ & 0.049$^{+0.186}_{-0.033}$ & 6.405$^{+0.003}_{-0.144}$ & 110.76 & 101 \\ 
\enddata
\tablecomments{*: For fitting purposes, the energy and the width of Fe K$\alpha$ line were limited to the range of 6.3-6.5\,keV and 0.001-0.25\,keV, respectively.\\
}
\end{deluxetable*}
\end{longrotatetable}

\bibliography{main}{}
\bibliographystyle{aasjournal}
\end{document}